\documentclass[10pt]{iopart}

\usepackage{iopams} 
\usepackage{array} 
\usepackage{graphicx}
\usepackage{units,hyperref}
\begin{document}

\title[Spin-orbit torque on nuclear spins exerted by a spin accumulation via hyperfine interactions]{Spin-orbit torque on nuclear spins exerted by a spin accumulation via hyperfine interactions}

\author{Adam B. Cahaya$^{1,2}$, Alejandro O. Leon$^3$ and Mohammad H. Fauzi$^{2,4}$}
\ead{adam@sci.ui.ac.id}
\address{$^1$Department of Physics, Faculty of Mathematics and Natural Sciences, Universitas Indonesia, Depok 16424, Indonesia}
\address{$^2$Research Center for Quantum Physics, National Research and Innovation Agency, South Tangerang, Banten, 15314 Indonesia}
\address{$^3$Departamento de F\'isica, Facultad de Ciencias Naturales,
Matem\'atica y del Medio Ambiente,
Universidad Tecnol\'ogica Metropolitana, Las Palmeras
3360, Ñuñoa 780-0003, Santiago, Chile}
\address{$^4$Research Collaboration Center for Quantum Technology 2.0, Bandung 40132, Indonesia}

\vspace{10pt}
\begin{indented}
\item[]
\end{indented}

\begin{abstract}

Spin-transfer and spin-orbit torques allow controlling magnetic degrees of freedom in various materials and devices. However, while the transfer of angular momenta between electrons has been widely studied, the contribution of nuclear spins has yet to be explored further. This article demonstrates that the hyperfine {coupling}, which consists of Fermi contact and dipolar interactions, can mediate the application of spin-orbit torques acting on nuclear spins. Our starting point is a sizable nuclear spin in a metal with electronic spin accumulation. Then, via the hyperfine interactions, the nuclear spin {modifies the} an electronic spin density. The reactions to the equilibrium and nonequilibrium components of the spin density is a torque on the nucleus with field-like and damping-like components, respectively. This \textit{nuclear spin-orbit torque} is a step toward stabilizing and controlling nuclear magnetic momenta, in magnitude and direction, and realizing nuclear spintronics.
\end{abstract}

%
\ioptwocol

\section{Introduction}

Nuclear magnetic resonance (NMR) is a powerful tool for investigating the physical and chemical properties of matter. However, nuclear magnetic momenta are three orders of magnitude smaller than the electronic ones, leading to weak nuclear polarization even at low temperatures and under high magnetic fields, hindering the NMR detection sensitivity~\cite{Hore2015-ti}. Several methods have been developed to overcome this limitation via generating a dynamic nuclear spin polarization, which can significantly improve the detection limit and resolution of NMR spectroscopy~\cite{Reimer20103}. These methods include optical~\cite{Tycko1996} and electrical techniques~\cite{Johnson2000,HIRAYAMA201168,Yang2018,PhysRevLett.125.106802}, and are particularly useful for studying low-dimensional materials, where the number of nuclei is too low for conventional NMR techniques to work~\cite{QUANTUMHALL}. The ongoing developments of new techniques for inducing nonequilibrium nuclear spin polarization, along with new detection schemes, are crucial for advancing the capabilities of NMR spectroscopy and expanding its potential applications~\cite{MLYNARIK20174}.

The techniques above employ angular-momentum transfer from the electron to the nuclear spin via the contact hyperfine interaction, and they have shown remarkable success in semiconductors with an s-wave conduction band, such as GaAs~\cite{Fauzi2022}. While both optical and electrical pumping and detection methods can achieve high levels of nuclear spin polarization and sensitivity {\cite{Costache,PhysRevB.77.125307,PhysRevB.77.085302,PhysRevLett.99.096804,ChapterSpinElectronics}}, the electrical one offers greater spatial selectivity and holds promise for integration with spintronic devices~\cite{Lin2017,Lin2019}.  Studies have demonstrated that nuclear spins in highly-doped GaAs semiconductors can be dynamically polarized by injecting spin-polarized electrons from an adjacent ferromagnetic contact~\cite{StrandPRL2003, StrandAPL2003, Salis2009}, and coherent manipulation of these polarized nuclear spins has also been achieved~\cite{Yamamoto2015}. However, these studies are primarily limited to GaAs semiconductors, and the possibility of observing similar phenomena in other platforms systems, such as magnetic multilayer devices has yet to be explored, both theoretically and experimentally. This motivates the quest for a spin-transfer-driven manipulation of the size and direction of nuclear spins in spintronic devices, such as magnetic heterostructures or metals with spin accumulation, \textit{i.e.,} the search for a \textit{nuclear spin-orbit torque} (NSOT).

\begin{figure}[t]
\centering
\includegraphics[width=\columnwidth]{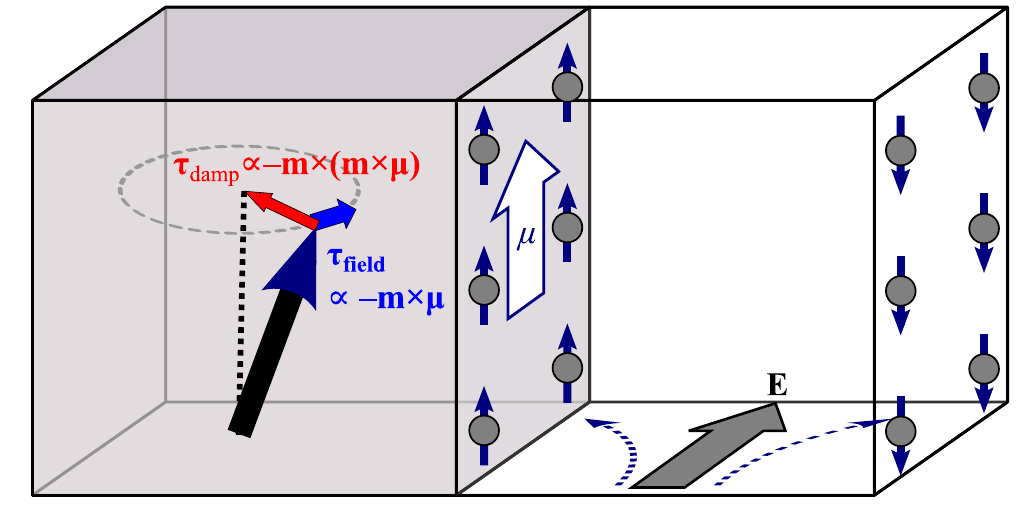}
\caption{Schematic setup to study spin-orbit torques. An electric field $\mathbf{E}$ generates a charge current in which, due to the spin-orbit interaction of the heavy metal {(lighter region)}, electrons with up and down spins separate, \textit{e.g.,} the spin-Hall effect~\cite{doi:10.1088/1468-6996/9/1/014105,PhysRevApplied.8.064023} takes place. As a result, an interfacial spin accumulation $\boldsymbol{\mu}$ is generated, exerting torque on the adjacent magnetization{, with unit vector $\mathbf{m}$, of a magnetic insulator (darker region). The spin-orbit torque has field- $\boldsymbol{\tau}_\textrm{field}\propto-\mathbf{m}\times\boldsymbol{\mu}$ and damping-like $\boldsymbol{\tau}_\textrm{damp}\propto-\mathbf{m}\times(\mathbf{m}\times\boldsymbol{\mu}) $ contributions.}}
\label{Fig1}
\end{figure}

In magnetic heterostructures, the exchange interaction between the magnetization and conduction electrons can generate a \textit{spin-transfer torque}~\cite{doi:10.1142/9789814273060_0008,Ralph2008,SLONCZEWSKI1996L1,PhysRevB.54.9353}. When the charge current is perpendicular to the magnetic layers, the spin-transfer torque relies on the spin polarization of the current, which is accomplished by adding a thicker magnet to the stack. On the other hand, \textit{spin-orbit torques}~\cite{doi:10.1088/1468-6996/9/1/014105,9427163} are exerted by a nonequilibrium spin accumulation at the interface with a heavy metal, as shown in figure~\ref{Fig1}. The spin-orbit coupling of the heavy metal induces the spin accumulation from a charge current {via spin-Hall effect~\cite{doi:10.1088/1468-6996/9/1/014105,PhysRevApplied.8.064023}. Consequently, the magnitude of the spin accumulation is ruled by the spin-Hall angle of the metal}. Depending on the current direction, spin-transfer and spin-orbit torques can enhance or compensate for the magnetic damping, allowing magnetization switching, inducing and moving magnetization textures, and generating permanent magnetization dynamics (see~\cite{Ralph2008,9427163} and references therein).
Beyond the bulk spin-orbit interaction of heavy metals, spin-orbit torques are driven by, or have contributions from, the Rashba-Edelstein effect~\cite{PhysRevB.104.184412,Auvray2018,PhysRevLett.117.116602,doi:10.1063/1.4990113} and emerge in two-dimensional materials~\cite{advs.202100847}. Additionally, spin-orbit torques can be manipulated by the crystal symmetries~\cite{MacNeill2016}.

The spin-mixing conductance is the parameter ruling the strength of the spin-orbit torque~\cite{10.1063/5.0024019,PhysRevResearch.3.013042}, as well as the spin-transfer torque and its reciprocal effect, namely, spin pumping~\cite{PhysRevB.66.224403,Jia11,weiler13}. Studies of the spin-mixing conductance at \textit{ferromagnet}$\vert$\textit{nonmagnetic metal} bilayers {indicate that it is affected by the interactions between spins and their environment at the interface}, such as crystal fields~\cite{PhysRevB.96.144434}, spin-orbit coupling~\cite{PhysRevLett.114.126602}, electric field screening~\cite{PhysRevB.103.094420}, gate voltage~\cite{PhysRevApplied.11.044060}, and orbital hybridization~\cite{PhysRevB.105.214438}. A recent study demonstrated the spin-current pumping from nuclear spin waves~\cite{Shiomi2018,Kikkawa2021}. In such a case, the spin transfer is mediated by the hyperfine {coupling~\cite{Cahaya2021}, which consists of dipole-dipole~\cite{10.1143/PTP.101.11,HIRAYAMA201168} and Fermi contact interactions~\cite{book:1567088}.}

In this article, we predict a nuclear spin-orbit torque from a spin accumulation. Our model consists of a stable nuclear spin $\mathbf{I}$, that we model as a classical vector, embedded in a free electron gas with a uniform spin accumulation $\boldsymbol{\mu}$. The interaction between the nuclear and electronic spins is the hyperfine coupling. While {the interactions between the nuclear spin and its environment, such as the Zeeman and quadrupolar couplings, and indirect exchange, may alter the magnetization equilibria or dynamics of a nuclear ensemble, our aim is to predict the nuclear spin-orbit torque, and such a derivation only requires the stable nuclear spin, the spin accumulation, and the hyperfine interaction between them. } The rest of this article is organized as follows. In section~\ref{Sec:LinearResponse}, we obtain the spin density of the electron gas. In section~\ref{Sec.Hyperfine}, we study the torque exerted by the electronic spin density on the nuclear spin. Section~\ref{Sec.Conclusion} summarizes our results. Finally, the detailed mathematical derivation of our results is in the supplementary material.

\section{Conduction electron spin density in the presence of a nuclear spin}
\label{Sec:LinearResponse}

Let us consider a uniform spin accumulation $\boldsymbol{\mu }$,  the electronic spin operator ${\mathbf{s}}=\hbar\boldsymbol{\sigma}/2$, and the following electron-gas Hamiltonian~\cite{PhysRevB.103.094420}
\begin{equation}
H=\frac{\mathbf{\hat{p}}^2}{2m_e} -\frac{1}{\hbar}{\mathbf{s}}\cdot\boldsymbol{\mu },
\end{equation}
where $\hbar$ is the reduced Plank constant, $\boldsymbol{\sigma}$ is the vector Pauli matrices, $m_e$ is the electron mass at rest, and $\mathbf{\hat{p}}$ is the momentum operator. We employ a free electron gas model with plane waves parametrized by their wavevector $\mathbf{p}$ (eigenvalue of $\mathbf{\hat{p}}/\hbar$). The conduction-electron magnetization $\mathbf{M}$ is related to the ensemble expectation value of the electronic spin, $\langle \mathbf{s}\rangle$, via $\mathbf{M}=-e\langle \mathbf{s}\rangle/m_e$, where $e$ is the modulus of the electron charge. The interaction between $\mathbf{M}$ and the net magnetic-type field acting on it, $\mathbf{B}$, is given by the Zeeman Hamiltonian
\begin{equation*}
H_{\rm Z}=-\int d^3r \mathbf{M}(\mathbf{r})\cdot \mathbf{B}(\mathbf{r}), 
\end{equation*}
where $\mathbf{r}$ is the position vector. The effect of $H_{\rm Z}$ on $\mathbf{M}(\mathbf{r})$ is studied using linear response theory~\cite{Kubo,PhysRevA.10.2461}. The $a-$th component of $\mathbf{M}$, $M_{a}$, may be written in terms of spin susceptibility tensor $\chi_{ab}$~\cite{CAHAYA2022168874},
\begin{small}
\begin{equation*}
M_{a}(\mathbf{r},t)=M_{0a} +\frac{e^2}{m_e^2}\sum_b\int d^3r'dt' \chi_{ab}(\mathbf{r}-\mathbf{r}',t-t')B_b(\mathbf{r}',t'),
\end{equation*}
\end{small}
\begin{eqnarray}
\chi_{ab} (\mathbf{r}-\mathbf{r}',t-t')=\frac{i}{\hbar}\theta(t-t')\left<\left[s_a(\mathbf{r},t),s_b(\mathbf{r}',t')\right]\right>_0,
\end{eqnarray}
where $\theta(x)$ is the Heaviside step function, the brackets $\langle f\rangle_0$ stand for the ensemble mean value of $f$ when $\mathbf{B}=0$, and $[f,g]$ is the commutator between $f$ and $g$. In the steady state, $\mathbf{B}$, $\boldsymbol\mu$, and $\chi_{ab}$ are time-independent. The contribution to $\mathbf{M}$ from the spin-accumulation is (see \textit{supplementary material})
\begin{equation}
\mathbf{M}_0=- \frac{e k_F}{2\pi^2\hbar}\boldsymbol{\mu},
\end{equation}
where $k_F$ is Fermi wavenumber. In reciprocal space, the susceptibility tensor reads $\chi_{ab}(q)= \delta_{ab}\chi_1(q)+\varepsilon_{abc}\mu _c\chi_2(q)$, where $q=\vert\mathbf{q}\vert$ is the modulus of the wavevector $\mathbf{q}$, $\mu_c$ is the $c-$th component of $\boldsymbol{\mu }$, and $\delta_{ab}$ and $\varepsilon_{abc}$ are the Kronecker delta and Levi-Civita symbol, respectively. The functions $\chi_1$ and $\chi_2$ are
\begin{eqnarray}
\chi_1(q)&=& \frac{m_ek_F}{8\pi^2} \left(1+\frac{k_F^2-\left(q/2\right)^2}{k_Fq}\ln\left|\frac{q+2k_F}{q-2k_F}\right|\right),\nonumber\\
\chi_2(q)&=&\frac{m_e^2}{8\pi\hbar^2}\frac{\theta(2k_F-q)}{q}, \label{Eq.chi2}
\end{eqnarray} 
where $f_\mathbf{p}=2\theta(k_F-p)$ is the Fermi-Dirac distribution of conduction electrons at zero temperature with energy $E_\mathbf{p}=\hbar^2p^2/(2m_e)$. In terms of $\chi_1$ and $\chi_2$, the magnetization of conduction electrons reads
\begin{equation}
\mathbf{M}(\mathbf{r})=\mathbf{M}_0 +\frac{e^2}{m_e^2}\int \frac{d^3qe^{i\mathbf{q}\cdot\mathbf{r}}}{(2\pi)^3}\left[\chi_1(q)-\chi_2(q)\boldsymbol{\mu}\times\right]\mathbf{B}(\mathbf{q}).
\label{Eq.MS}
\end{equation}
\begin{figure}[b!]
\centering
\includegraphics[width=\columnwidth]{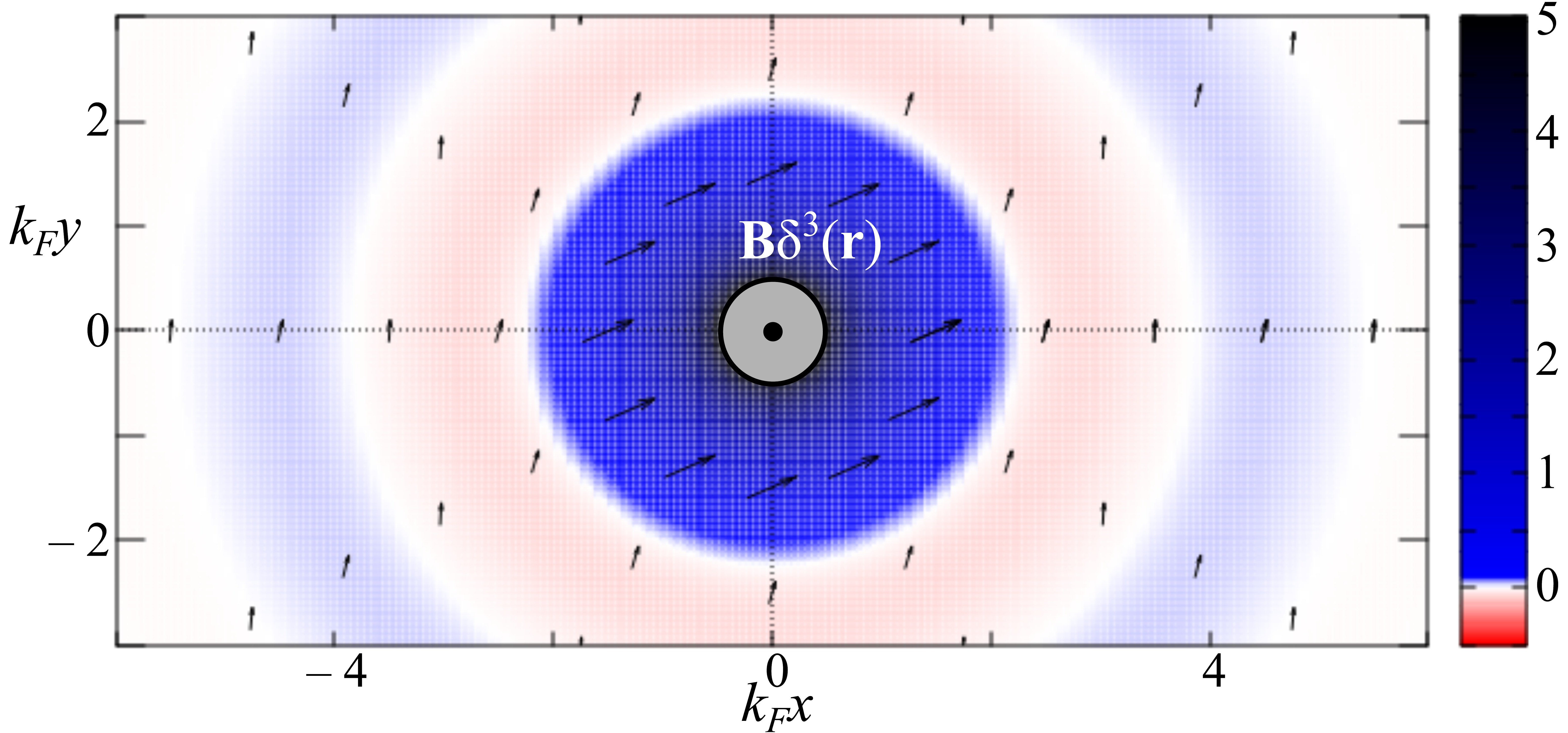}
\caption{Under a spin accumulation $\hat{\boldsymbol{\mu}}=\hat{\mathbf{y}}$, a localized magnetic field at the origin $\mathbf{B}(\mathbf{r})=B\delta^3(\mathbf{r})\hat{\mathbf{z}}$  induces a magnetization $\mathbf{M}$ (in arbitrary units). For illustrative purpose, the relative strength of $\hat{x}$, $\hat{y}$, and $\hat{z}$ components are exaggerated by choosing $\mu=\hbar^2k_F^2/(4m_e)$ and $B=20\hbar/(ek_F)$. Here, $\hat{x}$, $\hat{y}$, and $\hat{z}$ are unit vector along the $x$, $y$, and $z$ Cartesian axes, respectively.  $\mathbf{M}$ consists of three components which are parallel to the direction of $\hat{\boldsymbol{\mu}}=\hat{\mathbf{y}}$, $\hat{\mathbf{B}}=\hat{\mathbf{z}}$, and $\hat{\boldsymbol{\mu}}\times \hat{\mathbf{B}}=\hat{\mathbf{x}}$. Out-of-plane and in-plane components of $\mathbf{M}$ are indicated by a color map and arrows, respectively. The $z$-component of $\mathbf{M}$ diverges near the origin.}
\label{Fig.Magnetization}
\end{figure}
It is illustrative to evaluate the above expression for a static localized magnetic field, $\mathbf{B}\left(\mathbf{r}\right)=\mathbf{B_{\rm local}}\delta(\mathbf{r})$. In this case, the magnetization is the sum of three terms: $\mathbf{M}_0+\mathbf{M}_\parallel+\mathbf{M}_\perp$, where $\mathbf{M}_\parallel\propto f^{(1)}_{\rm con}\left(k_Fr\right) \mathbf{B_{\rm local}}$ is collinear to the source $ \mathbf{B_{\rm local}}$, with
\begin{equation}
f^{(1)}_{\rm con}(x)= \frac{\sin (2x)-2x\cos\left(2x\right)}{8x^4}
\label{EqF1}
\end{equation}
being the Friedel or RKKY oscillation function~\cite{KITTEL19691} arising from $\chi_1$. This is the equilibrium spin polarization of an electron gas along the field direction, $\mathbf{B_{\rm local}}$. On the other hand, $\mathbf{M}_\perp\propto f^{(2)}_{\rm con}\left(k_Fr\right)\boldsymbol{\mu} \times \mathbf{B_{\rm local}}$ is the nonequilibrium polarization given by the following response function~\cite{PhysRevB.103.094420,Kim1999}.
\begin{equation}
f^{(2)}_{\rm con}(x)= \frac{1-\cos\left(2x\right)}{2x^2},
\label{EqF2}
\end{equation}
that arises from $\chi_2$.
Figure~\ref{Fig.Magnetization} illustrates the sum of the $\mathbf{M}_0$, $\mathbf{M}_\parallel$, and $\mathbf{M}_\perp$ components. As this figure shows, the densities $\mathbf{M}_\parallel$ and $\mathbf{M}_\perp$ oscillate radially with wavenumber $2k_F$. While the equilibrium density, $\mathbf{M}_\parallel$, diverges at $\mathbf{r}=0$ and dominates near the source, it decays as $r^{-3}$. On the other hand, $\mathbf{M}_\perp\propto r^{-2}$ for large distances.

\subsection{Hyperfine coupling to a nuclear spin}

\begin{figure}[t!]
\centering
\includegraphics[width=\columnwidth]{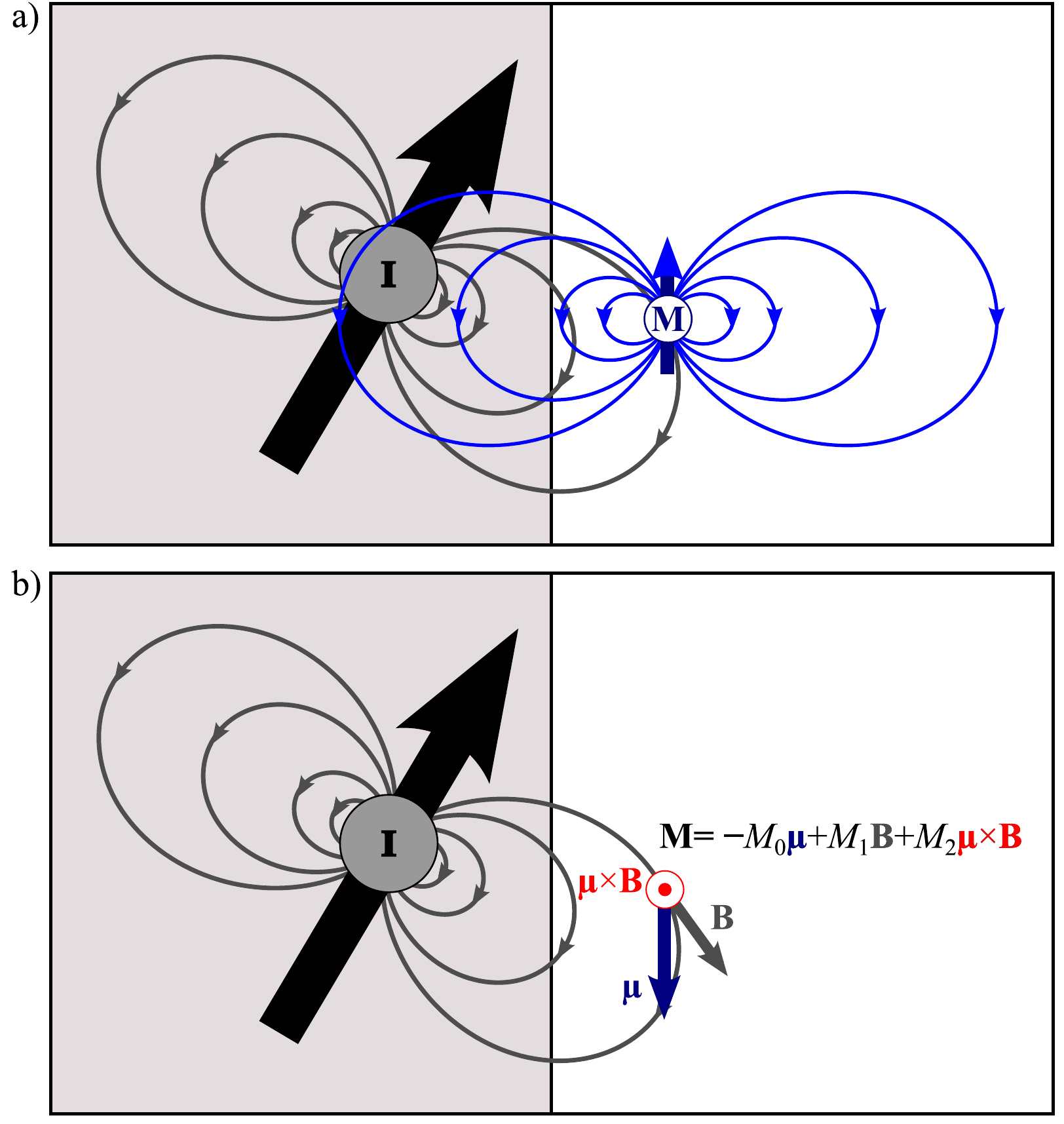}
\caption{(a) Dipole-dipole interaction between nuclear spin $\mathbf{I}$ and spin magnetic moments $\mathbf{M}$. (b) A spin accumulation $\boldsymbol\mu$ and a magnetic field $\mathbf{B}$ induce a conduction-electron magnetization $\mathbf{M}$ that consists of three components, parallel to the direction of $-\boldsymbol{\mu}$, $\mathbf{B}$, and $\boldsymbol{\mu}\times \mathbf{B}$.}
\label{Fig.Dipole}
\end{figure}

Let us consider a nuclear spin $\mathbf{I}$ at $\mathbf{r}=0$ that we model as a classical vector of magnitude $\vert\mathbf{I}\vert=\hbar I_p$. For a fully polarized nuclear spin, $I_p=\sqrt{I\left(I+1\right)}$ with $I$ being the nuclear spin quantum number. The magnetic hyperfine Hamiltonian~\cite{book:1567088} reads
\bigskip
\begin{eqnarray}
H_{\rm hf}=
-\frac{\mu_r\mu_0 \gamma_I}{4\pi}& & \int d^3r\left[\frac{8\pi\delta^3 \left(\mathbf{r}\right)}{3}\mathbf{M}\left(\mathbf{r}\right)\cdot \mathbf{I}
\nonumber\right.\\& &+\left.
\frac{3\left(\mathbf{M}\left(\mathbf{r}\right)\cdot\hat{\mathbf{r}}\right)\left(\mathbf{I}\cdot \hat{\mathbf{r}}\right)-\mathbf{M}\left(\mathbf{r}\right)\cdot\mathbf{I}}{r^3}\right],
\end{eqnarray}%
where $\gamma_I$ is the nuclear gyromagnetic ratio{;} $\mu_r$ and $\mu_0$ are the relative permeability and the permeability of free space{, respectively. The magnetic hyperfine interaction is responsible for spin wave excitation in nuclear ferromagnets~\cite{Suhl} and the nuclear spin relaxation via, \textit{e.g.,} the Korriga process~\cite{korringa1950nuclear,Vagner2003}. The latter effects arises from the electron spin fluctuations that we disregard here due to the spin accumulation.} The first term of $H_{\rm hf}$ is the Fermi contact coupling, while the second term is the dipolar interaction between nuclear and conduction-electron magnetic momenta, illustrated in figure~\ref{Fig.Dipole}a. Equating the Zeeman Hamiltonian $H_{\rm Z}$ to $H_{\rm hf}$, one identifies the following fields
\begin{enumerate}
\item Effective magnetic field from Fermi contact interaction, which localizes at $\mathbf{r}=0$,
\begin{eqnarray}
\mathbf{B}_{\rm con}(\mathbf{r})=\frac{2\mu_r\mu_0\gamma_I}{3} \delta^3 (\mathbf{r})\mathbf{I}.
\end{eqnarray}
\item Dipolar magnetic field, as illustrated in figure~\ref{Fig.Dipole}a
\begin{eqnarray}
\mathbf{B}_{\rm dip}(\mathbf{r})=\frac{\mu_r\mu_0\gamma_I}{4\pi} \frac{3\mathbf{r}\left(\mathbf{I}\cdot \mathbf{r}\right)-r^2\mathbf{I}}{r^5}.
\end{eqnarray}
\end{enumerate}
It is worth noting that $\mathbf{B}_{\rm con}(\mathbf{r})$ is equivalent to the exchange field by a local $3d$ or $4f$ magnetic moment. In the present case, it represents the coupling that emerges when the conduction-electron wavefunction reaches the nucleus~\cite{Kutzelnigg1988}. In reciprocal space,
\begin{eqnarray}
\mathbf{B}_{\rm con}(\mathbf{q})= \frac{2\mu_r\mu_0\gamma_I}{3}\mathbf{I}, \nonumber\\
\mathbf{B}_{\rm dip}(\mathbf{q})=\frac{\mu_r\mu_0\gamma_I}{3} \frac{q^2\mathbf{I}-3(\mathbf{I}\cdot \mathbf{q})\mathbf{q}}{q^2}.
\end{eqnarray}
Replacing these fields in equation~(\ref{Eq.MS}), we obtain the conduction-electron magnetization 
$\mathbf{M}(\mathbf{r})= \mathbf{M}_0+\mathbf{M}_{\rm con}(r)+\mathbf{M}_{\rm dip}(\mathbf{r})$,
where 
\begin{eqnarray*}
\mathbf{M}_{\rm con}(r)
&=&
\frac{e^2\mu_r\mu_0\gamma_Ik_F^4f^{(1)}_{\rm con}\left(k_Fr\right)}{6\pi^3m_e}\mathbf{I}\nonumber\\
& &+\frac{e^2\mu_r\mu_0\gamma_Ik_F^2f^{(2)}_{\rm con}\left(k_Fr\right)}{12\pi^3\hbar^2}{\mathbf{I}\times\boldsymbol{\mu }}
\end{eqnarray*}
is the magnetization due to Fermi contact field, shown in figure~\ref{Fig.Magnetization}, while the magnetization due to the dipolar magnetic field is
\begin{eqnarray*}
\mathbf{M}_{\rm dip}(\mathbf{r})
&=&\frac{e^2\mu_r\mu_0\gamma_Ik_F^4f^{(1)}_{\rm dip}\left(k_Fr\right)}{6\pi^3m_e}\left(3(\mathbf{I}\cdot\hat{\mathbf{r}})\hat{\mathbf{r}}-\mathbf{I}\right) \nonumber\\
& &+\frac{e^2\mu_r\mu_0\gamma_Ik_F^2f^{(2)}_{\rm dip}\left(k_Fr\right)}{12\pi^3\hbar^2}\boldsymbol{\mu}\times\left(3(\mathbf{I}\cdot\hat{\mathbf{r}})\hat{\mathbf{r}}-\mathbf{I}\right) ,
\end{eqnarray*}
and it is illustrated in figure~\ref{Fig.Dipole}b.
The radial functions are
\begin{eqnarray}
f^{(1)}_{\rm dip}(x)&=& \frac{x\cos\left(2x\right)-2\sin\left(2x\right)}{8x^4},\nonumber\\
f^{(2)}_{\rm dip}(x)&=& \frac{3\sin\left(2x\right)-2x\cos\left(2x\right)-4x}{8x^3},
\end{eqnarray}
\begin{figure}[t]
\centering
\includegraphics[width=\columnwidth]{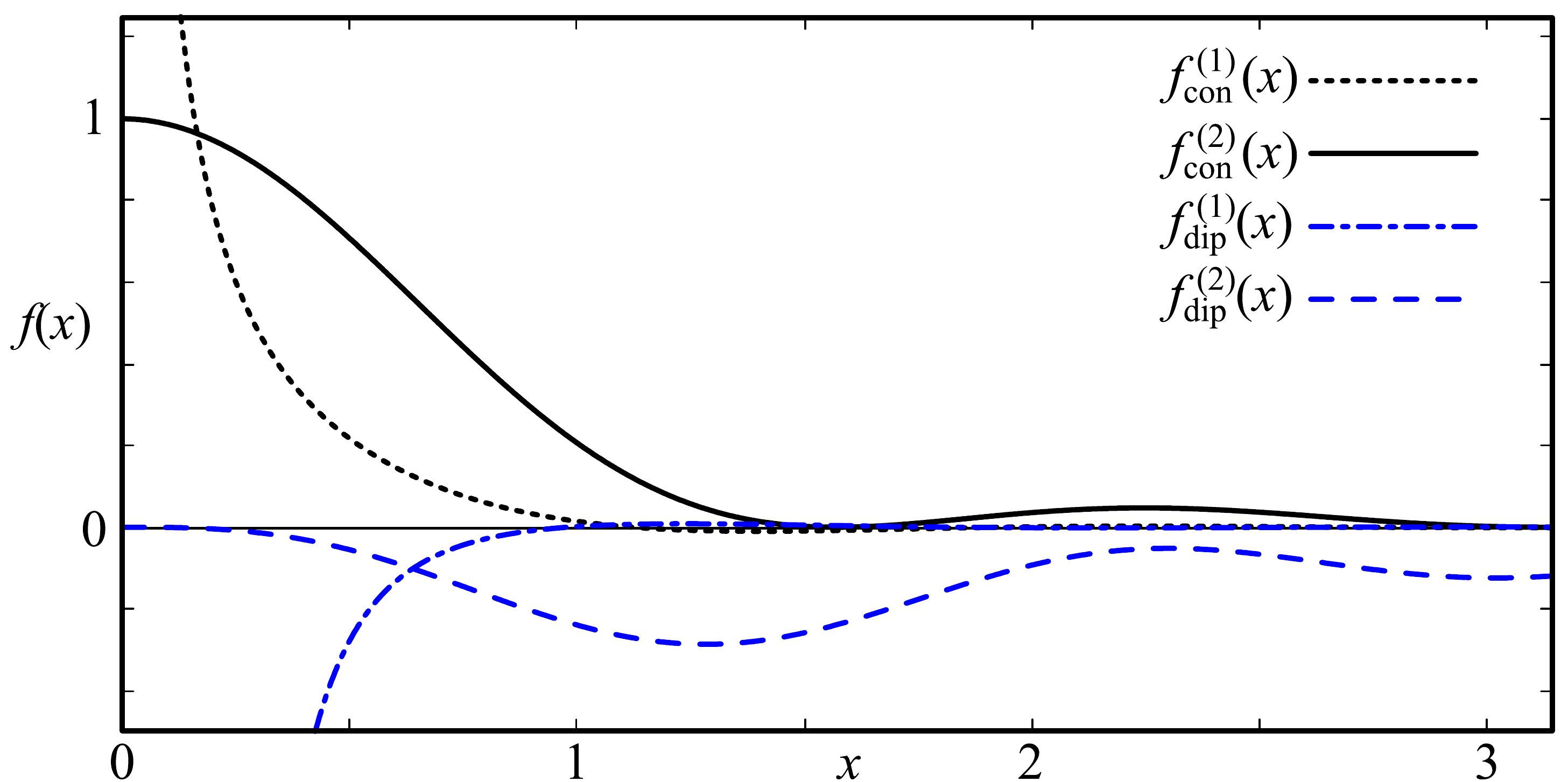}
\caption{Radial functions $f_{\rm con}^{(1)}$, $f_{\rm con}^{(2)}$ and $f_{\rm dip}^{(1)}$, $f_{\rm dip}^{(2)}$ associated with $\mathbf{M}_{\rm con}$ and $\mathbf{M}_{\rm dip}$.}
\label{Fig.functionradial}
\end{figure}
are illustrated in figure~\ref{Fig.functionradial} and are characterized by a $2k_F$ wavenumber and by decaying as $f^{(1)}_{\rm dip}\propto r^{-3}$ and $f^{(2)}_{\rm dip}\propto r^{-2}$ for $k_Fr\gg1$. Figure~\ref{Fig.functionradial} also shows the functions  $f_{\rm con}^{(1)}$ and $f_{\rm con}^{(2)}$, given by equations~(\ref{EqF1}) and~(\ref{EqF2}). The nonequilibrium part of the Fermi contact and dipolar magnetization, rules by $f_{\rm con}^{(2)}$ and $f^{(2)}_{\rm dip}$, respectively, dominate far from the nuclear spin.

Using the asymptotic values of the radial functions, the Taylor expansion of $\mathbf{M}$ for $r\ll k_F^{-1}$ reads
\begin{eqnarray}
\mathbf{M}(\mathbf{r})
\approx -\boldsymbol{\mu} \frac{e k_F}{2\pi^2\hbar}-\frac{e^2\mu_r\mu_0\gamma_Ik_F^2}{12\pi^3\hbar^2}\boldsymbol{\mu}\times\mathbf{I}\nonumber\\
+ \frac{e^2\mu_r\mu_0\gamma_Ik_F^4}{6\pi^3m_e }\left(\frac{\mathbf{I}}{3k_Fr} -3\frac{3(\mathbf{I}\cdot\hat{\mathbf{r}})\hat{\mathbf{r}}-\mathbf{I}}{8(k_Fr)^3}\right)
+\mathcal{O}\left(r^2\right).
\end{eqnarray}
Let us stress that terms proportional to $f^{(2)}_{\rm con}$ and $f^{(2)}_{\rm dip}$ in $\mathbf{M}_{\rm con}$ and $\mathbf{M}_{\rm dip}$, respectively, are nonequilibrium because they require the finite nonequilibrium spin accumulation $\boldsymbol{\mu }$.

\section{Nuclear spin-orbit torque due to magnetic hyperfine interaction}
\label{Sec.Hyperfine}

\begin{figure}[b]
\centering
\includegraphics[width=\columnwidth]{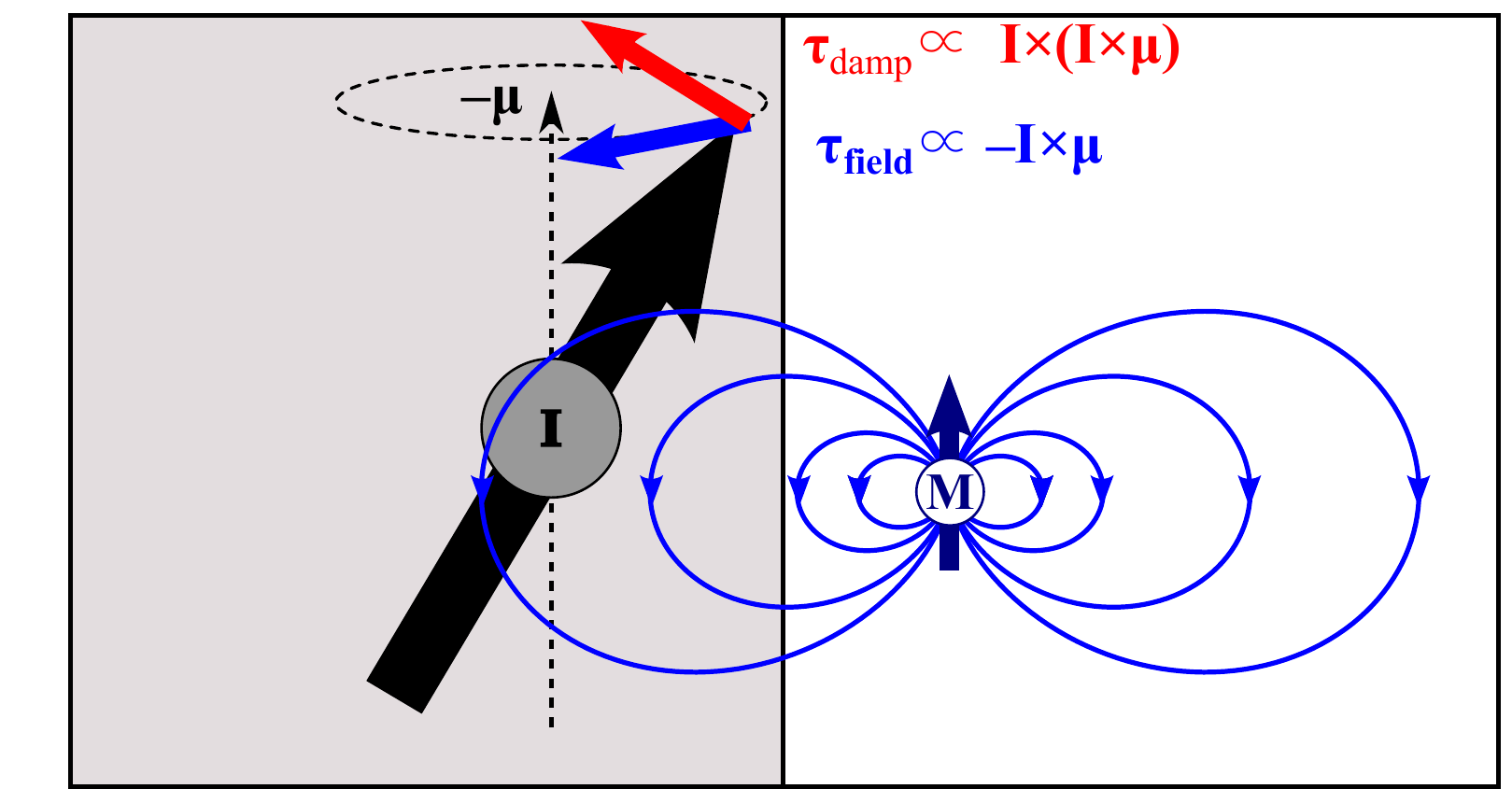}
\caption{The hyperfine interaction between the nucleus and the conduction-electron magnetization, $\mathbf{M}$, generates a torques with field-like $\boldsymbol{\tau}_\textrm{field}\propto-\mathbf{I}\times\boldsymbol{\mu}$ and damping-like $\boldsymbol{\tau}_\textrm{damp}\propto\mathbf{I}\times(\mathbf{I}\times\boldsymbol{\mu}) $, \textit{i.e., } a nuclear spin-orbit torque. }
\label{Fig.DipoleTorque}
\end{figure}

The spin-orbit torque acting on the nuclear spin, $\boldsymbol\tau_{\rm NSOT}=\left(d\mathbf{I}/dt\right)_{\rm NSOT}$, can be determined from $H_{\rm hf}$ 
\begin{equation}
\boldsymbol\tau_{\rm NSOT}=- \mathbf{I}\times \frac{\delta H_{\rm hf}}{\delta \mathbf{I}}. \label{Eq.IxB}
\end{equation}
Using $\mathbf{M}(\mathbf{r})= \mathbf{M}_0+\mathbf{M}_{\rm con}(r)+\mathbf{M}_{\rm dip}(\mathbf{r})$, we obtain the following torque per unit area (see Supplemental Material for the detailed derivation)
\begin{eqnarray}
\frac{\boldsymbol\tau_{\rm NSOT}}{a^2}&=&-g_{\rm field}\left(\mathbf{m}\times\boldsymbol{\mu }\right)+g_{\rm damp}\mathbf{m}\times\left(\mathbf{m}\times\boldsymbol{\mu }\right),\nonumber\\
g_{\rm field}&=&\frac{e\mu_r\mu_0\gamma_Ik_F I_p}{3\pi^2a^2},\nonumber\\
g_{\rm damp}&=&\frac{5(e\mu_r\mu_0\gamma_Ik_F)^2I_p^2}{72\pi^3a^2}, \label{Eq.NSOT}
\end{eqnarray}
where $a$ is the lattice constant, $\mathbf{m}=\mathbf{I}/\vert\mathbf{I}\vert$ and $g_{\rm nuclear}\equiv g_{\rm damp}+ig_{\rm field}$ is the nuclear version of complex-valued spin-mixing conductance. The dipole contribution is four times smaller than the the exchange-like Fermi contact field. 

Figure~\ref{Fig.DipoleTorque} illustrates the directions of field-like and damping-like torques that are induced by the hyperfine interaction. It is worth noting that depending on the nucleus, its magnetic momentum can be parallel ($\gamma_I>0$) or antiparallel ($\gamma_I<0$) to the nuclear spin, which changes the field-type-torque direction. Besides, the sign of the NSOT changes when reversing the direction of the spin accumulation, as it occurs with the (electronic) spin-orbit torque. Note that larger nuclear spins have a stronger influence on conduction electrons, as shown by the $\mathbf{M}_{\rm con},\mathbf{M}_{\rm dip}\propto \mathbf{I}$ proportionality. Therefore, the response to the conduction-electron magnetization, \textit{i.e.,} the NSOT, is characterized by $g_{\rm damp}\propto \vert\mathbf{I}\vert^2$. While our theory relies on a sizable nuclear spin that can be treated as a classical vector, the separation between the temporal scales of conduction electrons and the magnetization of an ensemble of nuclear spins allows us to conjecture that the NSOT may be added to the Bloch equation to obtain the full dynamics of the nuclear system.

Dependency of the torques on $\mu_r\mu_0$ indicates that NSOT is large for a bilayer of ferromagnet with high spin nucleus and nonmagnetic metal with high magnetic permeability, such as Fe ($\mu_r\mu_0\sim 0.25$ H/m~\cite{YENSEN1928503,Schenck1996}).
For $^{57}$Fe, $I=1/2$, $k_F\sim 2$ \AA$^{-1}$, lattice constant $a\sim 2$ \AA, we estimate that $g_{\rm field}$ is orders of magnitude larger than $g_{\rm damp}$
\begin{eqnarray*}
g_{\rm field}\sim 5\times 10^{15} \ {\rm m}^{-2},\\
g_{\rm damp}\sim 2\times 10^{12} \ {\rm m}^{-2}.
\end{eqnarray*} 
{Note that the field-type contribution dominates over the damping-like one, $g_{\rm field}\gg g_{\rm damp}$. This relationship is opposite to the one between the real and imaginary parts of the electronic spin-mixing conductance, in which the damping-like contribution is substantially larger. The value of $g_{\rm damp}$ is similar measurement of nuclear spin pumping in Ref~\cite{Shiomi2018}.}

{The set of Eqs.~(\ref{Eq.NSOT}) shows that the nuclear spin-orbit torque has the same mathematical form as the electronic spin-transfer and spin-orbit torques, suggesting that a nuclear spin ensemble can be manipulated using spin-orbit torques. Indeed, the $-g_{\rm field}\mathbf{m}\times\boldsymbol{\mu }$ torque is identical to the one exerted by a magnetic field and $g_{\rm damp}\mathbf{m}\times\left(\mathbf{m}\times\boldsymbol{\mu }\right)$ has the structure of a Landau-Lifshitz damping \cite{Bertotti2009} with a direction and sign determined by $\boldsymbol{\mu }$ and, therefore, controlled by the current. To exemplify the possible applications of the NSOT, consider a nuclear spin under a large net magnetic field $\mathbf{B}_{\rm ext}=B_0\hat{\mathbf{z}}$ and $\boldsymbol{\mu }= \mu\hat{\mathbf{z}}$ such that $\vert m_x\vert,\vert m_y\vert\ll m_z\approx1$, where $m_j=\mathbf{m}\cdot\mathbf{e_j}$. Then, in the $x-y$ plane, the $j-$th component of the NSOT reads
\begin{eqnarray}
\frac{dI_j}{dt}=& \tau_{\rm damp}=g_\mathrm{damp}a^2\left[\mathbf{m}\times\left(\mathbf{m}\times\boldsymbol{\mu }\right)\right]_j\nonumber\\
\frac{dm_j}{dt}=& \frac{g_{\rm damp} a^2\mu m_j}{\hbar I_p},
\end{eqnarray}
and it can cancel the transverse relaxation term of the Bloch equation \cite{Kittel2018} for $\mu\geq \left(g_{\rm damp} a^2T_2\right)^{-1}\hbar I_p$, where $T_2$ is the transverse relaxation time. As described in the Supplemental Material, for solid state quantum devices with $T_2\sim 10^{-3}$~s \cite{Yusa2005,Ota2007}, the critical value of the spin accumulation is in the order of $7\times10^{-6}$~eV, which is the same order of magnitude as the one reported in the literature~\cite{Liu_2018}. This magnitude of  can be achieved for charge-current density
$j\geq\frac{\mu \sigma }{2e\lambda \theta}\sim 2\times 10^{11} \textrm{ A/m}^2,$
where $\sigma\sim 10^7 \ \Omega^{-1}$m$^{-1}$, $\theta\sim 0.1$, $\lambda\sim 2$~nm are the conductivity, spin Hall angle, and spin-diffusion length, respectively. It is worth noting that the spin-Hall angle, $\theta$ is the parameter that rules the efficiency of the spin-Hall effect responsible for the spin accumulation $\boldsymbol{\mu}$. While the resulting magnetization dynamics in the above scenario may depend on several physical ingredients, including the Zeeman, dipolar, and quadrupolar couplings, as well as indirect spin-spin exchange, the damping compensation promises the possibility of amplifying the transverse magnetization components and potentially switching nuclear spins or generating permanent dynamics.

Finally, let us stress that our main assumption is the existence of a stable nuclear spin that can be modeled by a classical vector. The time evolution of this vector, or an ensemble of nuclear spins, is given by the appropriate Bloch equation \cite{Kittel2018} with several torques accounting for the environment of the nuclear spins.}

\section{Conclusions and remarks}
\label{Sec.Conclusion}
This work predicts a torque acting on nuclear spins embedded in an electronic spin accumulation, namely a \textit{nuclear spin-orbit torque} (NSOT). Our model consist of a sizable nuclear spin in a simple metal described by plane waves. {represented by a classical vector in a simple metal described by plane waves. The use of a classical vector for the spin assumes a spin polarization given by, \textit{e.g.,} a large magnetic field or the exchange coupling with atomic (3d) magnetic moments. Consequently, it is worth noting that our results are independent of the interactions between the nuclear spin and its the environment, such as the Zeeman, dipolar, quadrupolar couplings, indirect exchange with other nuclei, interactions with the lattice, as far as the model assumptions are satisfied.}

By using linear response theory, we show that the magnetic hyperfine interaction induces a magnetization on conduction electrons around the nuclear spin. When the metal has spin accumulation, the electronic magnetization is not parallel to the nuclear spin and exerts a NSOT on the nucleus.

The NSOT has field-like and damping-like components, characterized by $g_{\rm field}$ and $g_{\rm damp}$, respectively. $g_{\rm field}$ and $g_{\rm damp}$ are analogous to the imaginary and real parts of complex-valued spin-mixing conductance which describes spin-transfer and spin-orbit torques in electronic systems. Furthermore, $g_{\rm damp}$ comes from the nonequilibrium part of the response function, $\chi_2$, which is also responsible for damping-like spin-transfer torque~\cite{PhysRevB.103.094420} and spin pumping~\cite{PhysRevB.96.144434}. Contribution of dipolar interaction in $g_{\rm damp}$ is smaller than Fermi contact, in accordance with \cite{Cahaya2021}. The value of the total $g_{\rm damp}$ is in agreement with observation by \cite{Shiomi2018}.

Dependency of the torques on $I$ and $\mu_r\mu_0$ indicates that the NSOT is large for a bilayer of ferromagnet with high spin nucleus and nonmagnetic metal with high magnetic permeability, such as Fe, which has a naturally occurring and stable isotope $^{57}$Fe. { In NMR quantum computing, nuclear spin is a candidate for qubit. Since qubit has the potential to the miniaturization of memory, the application of nuclear spin by means of {NSOT} could} lead to further miniaturization of spintronics devices.

\ack
ABC acknowledges the support of PUTI Grant No. NKB-1448/UN2.RST/HKP.05.00/2022. AOL acknowledges financial support in Chile from ANID FONDECYT 11230120, and ANID FONDECYT 1210353.

\section*{References}

\begin{thebibliography}{10}
\expandafter\ifx\csname url\endcsname\relax
  \def\url#1{{\tt #1}}\fi
\expandafter\ifx\csname urlprefix\endcsname\relax\def\urlprefix{URL }\fi
\providecommand{\eprint}[2][]{\url{#2}}

\bibitem{Hore2015-ti}
Hore P 2015 {\em Nuclear Magnetic Resonance\/} 2nd ed Oxford Chemistry Primers
  (London, England: Oxford University Press)

\bibitem{Reimer20103}
Reimer J~A 2010 Nuclear hyperpolarization in solids and the prospects for
  nuclear spintronics {\em Solid State Nuclear Magnetic Resonance\/} {\bf 37}
  3--12 ISSN 0926-2040
  \urlprefix\url{https://doi.org/10.1016/j.ssnmr.2010.04.001}

\bibitem{Tycko1996}
Tycko R and Reimer J~A 1996 Optical pumping in solid state nuclear magnetic
  resonance {\em The Journal of Physical Chemistry\/} {\bf 100} 13240--13250
  ISSN 0022-3654 \urlprefix\url{https://doi.org/10.1021/jp953667u}

\bibitem{Johnson2000}
Johnson M 2000 Dynamic nuclear polarization by spin injection {\em Applied
  Physics Letters\/} {\bf 77} 1680--1682 (\textit{Preprint}
  \eprint{https://doi.org/10.1063/1.1310173})
  \urlprefix\url{https://doi.org/10.1063/1.1310173}

\bibitem{HIRAYAMA201168}
Hirayama Y 2011 2.03 - contact hyperfine interactions in semiconductor
  heterostructures {\em Comprehensive Semiconductor Science and Technology\/}
  ed Bhattacharya P, Fornari R and Kamimura H (Amsterdam: Elsevier) pp 68--94
  ISBN 978-0-444-53153-7
  \urlprefix\url{https://doi.org/10.1016/B978-0-44-453153-7.00081-X}

\bibitem{Yang2018}
Yang K, Willke P, Bae Y, Ferr{\'{o}}n A, Lado J~L, Ardavan A,
  Fern{\'{a}}ndez-Rossier J, Heinrich A~J and Lutz C~P 2018 Electrically
  controlled nuclear polarization of individual atoms {\em Nature
  Nanotechnology\/} {\bf 13} 1120--1125
  \urlprefix\url{https://doi.org/10.1038/s41565-018-0296-7}

\bibitem{PhysRevLett.125.106802}
Jiang Z, Soghomonian V and Heremans J~J 2020 Dynamic nuclear spin polarization
  induced by the edelstein effect at bi(111) surfaces {\em Phys. Rev. Lett.\/}
  {\bf 125}(10) 106802
  \urlprefix\url{https://doi.org/10.1103/PhysRevLett.125.106802}

\bibitem{QUANTUMHALL}
Hirayama Y 2013 {\em Hyperfine interactions in quantum Hall regime\/} (World
  Scientific) pp 730--753
  \urlprefix\url{https://doi.org/10.1142/9789814360760\_0038}

\bibitem{MLYNARIK20174}
Mlynárik V 2017 Introduction to nuclear magnetic resonance {\em Analytical
  Biochemistry\/} {\bf 529} 4--9 ISSN 0003-2697 introduction to in vivo
  Magnetic Resonance Spectroscopy (MRS): A method to non-invasively study
  metabolism \urlprefix\url{https://doi.org/10.1016/j.ab.2016.05.006}

\bibitem{Fauzi2022}
Fauzi M~H and Hirayama Y 2022 Hyperfine-mediated transport in a one-dimensional
  channel {\em Quantum Hybrid Electronics and Materials\/} ed Hirayama Y,
  Hirakawa K and Yamaguchi H (Singapore: Springer Nature Singapore) pp 257--276
  ISBN 978-981-19-1201-6

\bibitem{Costache}
Costache M~V and Valenzuela S~O 2010 Experimental spin ratchet {\em Science\/}
  {\bf 330} 1645--1648 (\textit{Preprint}
  \eprint{https://www.science.org/doi/pdf/10.1126/science.1196228})
  \urlprefix\url{https://www.science.org/doi/abs/10.1126/science.1196228}

\bibitem{PhysRevB.77.125307}
Makhonin M~N, Tartakovskii A~I, Van'kov A~B, Drouzas I, Wright T,
  Skiba-Szymanska J, Russell A, Fal'ko V~I, Skolnick M~S, Liu H~Y and Hopkinson
  M 2008 Long nuclear spin polarization decay times controlled by optical
  pumping in individual quantum dots {\em Phys. Rev. B\/} {\bf 77}(12) 125307
  \urlprefix\url{https://link.aps.org/doi/10.1103/PhysRevB.77.125307}

\bibitem{PhysRevB.77.085302}
Yang W and Liu R~B 2008 Decoherence of coupled electron spins via nuclear spin
  dynamics in quantum dots {\em Phys. Rev. B\/} {\bf 77}(8) 085302
  \urlprefix\url{https://link.aps.org/doi/10.1103/PhysRevB.77.085302}

\bibitem{PhysRevLett.99.096804}
Baugh J, Kitamura Y, Ono K and Tarucha S 2007 Large nuclear overhauser fields
  detected in vertically coupled double quantum dots {\em Phys. Rev. Lett.\/}
  {\bf 99}(9) 096804
  \urlprefix\url{https://link.aps.org/doi/10.1103/PhysRevLett.99.096804}

\bibitem{ChapterSpinElectronics}
Tarucha S, Stopa M, Sasaki S and Ono K 2006 {Probing and manipulating spin
  effects in quantum dots} {\em {Concepts in Spin Electronics}\/} ed Maekawa S
  (Oxford University Press) ISBN 9780198568216 (\textit{Preprint}
  \eprint{https://academic.oup.com/book/0/chapter/149757956/chapter-ag-pdf/44985080/book\_6173\_section\_149757956.ag.pdf})
  \urlprefix\url{https://doi.org/10.1093/acprof:oso/9780198568216.003.0003}

\bibitem{Lin2017}
Lin Z, Rasly M and Uemura T 2017 Electrical detection of nuclear spin-echo
  signals in an electron spin injection system {\em Applied Physics Letters\/}
  {\bf 110} 232404 (\textit{Preprint}
  \eprint{https://doi.org/10.1063/1.4985650})
  \urlprefix\url{https://doi.org/10.1063/1.4985650}

\bibitem{Lin2019}
Lin Z, Pan D, Rasly M and Uemura T 2019 Electrical spin injection into
  algaas/gaas-based two-dimensional electron gas systems with co2mnsi spin
  source up to room temperature {\em Applied Physics Letters\/} {\bf 114}
  012405 (\textit{Preprint} \eprint{https://doi.org/10.1063/1.5077027})
  \urlprefix\url{https://doi.org/10.1063/1.5077027}

\bibitem{StrandPRL2003}
Strand J, Schultz B~D, Isakovic A~F, Palmstr\o{}m C~J and Crowell P~A 2003
  Dynamic nuclear polarization by electrical spin injection in
  ferromagnet-semiconductor heterostructures {\em Phys. Rev. Lett.\/} {\bf
  91}(3) 036602 \urlprefix\url{https://doi.org/10.1103/PhysRevLett.91.036602}

\bibitem{StrandAPL2003}
Strand J, Isakovic A~F, Lou X, Crowell P~A, Schultz B~D and Palmstrøm C~J 2003
  Nuclear magnetic resonance in a ferromagnet–semiconductor heterostructure
  {\em Applied Physics Letters\/} {\bf 83} 3335--3337 (\textit{Preprint}
  \eprint{https://doi.org/10.1063/1.1620685})
  \urlprefix\url{https://doi.org/10.1063/1.1620685}

\bibitem{Salis2009}
Salis G, Fuhrer A and Alvarado S~F 2009 Signatures of dynamically polarized
  nuclear spins in all-electrical lateral spin transport devices {\em Phys.
  Rev. B\/} {\bf 80}(11) 115332
  \urlprefix\url{https://doi.org/10.1103/PhysRevB.80.115332}

\bibitem{Yamamoto2015}
Yamamoto A, Ando Y, Shinjo T, Uemura T and Shiraishi M 2015 Spin transport and
  spin conversion in compound semiconductor with non-negligible spin-orbit
  interaction {\em Phys. Rev. B\/} {\bf 91}(2) 024417
  \urlprefix\url{https://doi.org/10.1103/PhysRevB.91.024417}

\bibitem{doi:10.1142/9789814273060_0008}
Baraduc C, Chshiev M and Ebels U 2010 {\em Introduction to spin transfer
  torque\/} (World Scientific) pp 173--192
  \urlprefix\url{https://doi.org/10.1142/9789814273060}

\bibitem{Ralph2008}
Ralph D and Stiles M 2008 Spin transfer torques {\em Journal of Magnetism and
  Magnetic Materials\/} {\bf 320} 1190--1216
  \urlprefix\url{https://doi.org/10.1016/j.jmmm.2007.12.019}

\bibitem{SLONCZEWSKI1996L1}
Slonczewski J 1996 Current-driven excitation of magnetic multilayers {\em
  Journal of Magnetism and Magnetic Materials\/} {\bf 159} L1--L7 ISSN
  0304-8853 \urlprefix\url{https://doi.org/10.1016/0304-8853(96)00062-5}

\bibitem{PhysRevB.54.9353}
Berger L 1996 Emission of spin waves by a magnetic multilayer traversed by a
  current {\em Phys. Rev. B\/} {\bf 54}(13) 9353--9358
  \urlprefix\url{https://doi.org/10.1103/PhysRevB.54.9353}

\bibitem{doi:10.1088/1468-6996/9/1/014105}
Takahashi S and Maekawa S 2008 Spin current, spin accumulation and spin hall
  effect {\em Science and Technology of Advanced Materials\/} {\bf 9} 014105
  pMID: 27877931 \urlprefix\url{https://doi.org/10.1088/1468-6996/9/1/014105}

\bibitem{9427163}
Shao Q, Li P, Liu L, Yang H, Fukami S, Razavi A, Wu H, Wang K, Freimuth F,
  Mokrousov Y, Stiles M~D, Emori S, Hoffmann A, Åkerman J, Roy K, Wang J~P,
  Yang S~H, Garello K and Zhang W 2021 Roadmap of spin–orbit torques {\em
  IEEE Transactions on Magnetics\/} {\bf 57} 1--39
  \urlprefix\url{https://doi.org/10.1109/TMAG.2021.3078583}

\bibitem{PhysRevApplied.8.064023}
Spiesser A, Saito H, Fujita Y, Yamada S, Hamaya K, Yuasa S and Jansen R 2017
  Giant spin accumulation in silicon nonlocal spin-transport devices {\em Phys.
  Rev. Applied\/} {\bf 8}(6) 064023
  \urlprefix\url{https://doi.org/10.1103/PhysRevApplied.8.064023}

\bibitem{PhysRevB.104.184412}
Lee W~B, Kim S~B, Kim K~W, Lee K~J, Koo H~C and Choi G~M 2021 Direct
  observation of spin accumulation and spin-orbit torque driven by
  rashba-edelstein effect in an inas quantum-well layer {\em Phys. Rev. B\/}
  {\bf 104}(18) 184412
  \urlprefix\url{https://doi.org/10.1103/PhysRevB.104.184412}

\bibitem{Auvray2018}
Auvray F, Puebla J, Xu M, Rana B, Hashizume D and Otani Y 2018 Spin
  accumulation at nonmagnetic interface induced by direct
  rashba{\textendash}edelstein effect {\em Journal of Materials Science:
  Materials in Electronics\/} {\bf 29} 15664--15670
  \urlprefix\url{https://doi.org/10.1007/s10854-018-9162-5}

\bibitem{PhysRevLett.117.116602}
Nakayama H, Kanno Y, An H, Tashiro T, Haku S, Nomura A and Ando K 2016
  Rashba-edelstein magnetoresistance in metallic heterostructures {\em Phys.
  Rev. Lett.\/} {\bf 117}(11) 116602
  \urlprefix\url{https://doi.org/10.1103/PhysRevLett.117.116602}

\bibitem{doi:10.1063/1.4990113}
Puebla J, Auvray F, Xu M, Rana B, Albouy A, Tsai H, Kondou K, Tatara G and
  Otani Y 2017 Direct optical observation of spin accumulation at nonmagnetic
  metal/oxide interface {\em Applied Physics Letters\/} {\bf 111} 092402
  \urlprefix\url{https://doi.org/10.1063/1.4990113}

\bibitem{advs.202100847}
Tang W, Liu H, Li Z, Pan A and Zeng Y~J 2021 Spin-orbit torque in van der
  waals-layered materials and heterostructures {\em Advanced Science\/} {\bf 8}
  2100847 \urlprefix\url{https://doi.org/10.1002/advs.202100847}

\bibitem{MacNeill2016}
MacNeill D, Stiehl G~M, Guimaraes M~H~D, Buhrman R~A, Park J and Ralph D~C 2016
  Control of spin{\textendash}orbit torques through crystal symmetry in
  {WTe}2/ferromagnet bilayers {\em Nature Physics\/} {\bf 13} 300--305
  \urlprefix\url{https://doi.org/10.1038/nphys3933}

\bibitem{10.1063/5.0024019}
Amin V~P, Haney P~M and Stiles M~D 2020 {Interfacial spin–orbit torques} {\em
  Journal of Applied Physics\/} {\bf 128} ISSN 0021-8979 151101
  \urlprefix\url{https://doi.org/10.1063/5.0024019}

\bibitem{PhysRevResearch.3.013042}
Hayashi H, Musha A, Sakimura H and Ando K 2021 Spin-orbit torques originating
  from the bulk and interface in pt-based structures {\em Phys. Rev. Res.\/}
  {\bf 3}(1) 013042
  \urlprefix\url{https://doi.org/10.1103/PhysRevResearch.3.013042}

\bibitem{PhysRevB.66.224403}
Tserkovnyak Y, Brataas A and Bauer G~E~W 2002 Spin pumping and magnetization
  dynamics in metallic multilayers {\em Phys. Rev. B\/} {\bf 66}(22) 224403
  \urlprefix\url{https://doi.org/10.1103/PhysRevB.66.224403}

\bibitem{Jia11}
Jia X, Liu K, Xia K and Bauer G 2011 Spin transfer torque on magnetic
  insulators {\em Europhys. Lett.\/} {\bf 96} 17005
  \urlprefix\url{https://doi.org/10.1209/0295-5075/96/17005}

\bibitem{weiler13}
Weiler M, Althammer M, Schreier M, Lotze J, Pernpeintner M, Meyer S, Huebl H,
  Gross R, Kamra A, Xiao J, Chen Y~T, Jiao H, Bauer G and Goennenwein S 2013
  Experimental test of the spin mixing interface conductivity concept {\em
  Phys. Rev. Lett.\/} {\bf 111}(17) 176601
  \urlprefix\url{https://doi.org/10.1103/PhysRevLett.111.176601}

\bibitem{PhysRevB.96.144434}
Cahaya A~B, Leon A~O and Bauer G~E~W 2017 Crystal field effects on spin pumping
  {\em Phys. Rev. B\/} {\bf 96}(14) 144434
  \urlprefix\url{https://doi.org/10.1103/PhysRevB.96.144434}

\bibitem{PhysRevLett.114.126602}
Chen K and Zhang S 2015 Spin pumping in the presence of spin-orbit coupling
  {\em Phys. Rev. Lett.\/} {\bf 114}(12) 126602
  \urlprefix\url{https://doi.org/10.1103/PhysRevLett.114.126602}

\bibitem{PhysRevB.103.094420}
Cahaya A~B and Majidi M~A 2021 Effects of screened coulomb interaction on spin
  transfer torque {\em Phys. Rev. B\/} {\bf 103}(9) 094420
  \urlprefix\url{https://doi.org/10.1103/PhysRevB.103.094420}

\bibitem{PhysRevApplied.11.044060}
Wang L, Lu Z, Xue J, Shi P, Tian Y, Chen Y, Yan S, Bai L and Harder M 2019
  Electrical control of spin-mixing conductance in a
  ${\mathrm{y}}_{3}{\mathrm{fe}}_{5}{\mathrm{o}}_{12}/$platinum bilayer {\em
  Phys. Rev. Applied\/} {\bf 11}(4) 044060
  \urlprefix\url{https://doi.org/10.1103/PhysRevApplied.11.044060}

\bibitem{PhysRevB.105.214438}
Cahaya A~B, Sitorus R~M, Azhar A, Nugraha A~R~T and Majidi M~A 2022 Enhancement
  of spin-mixing conductance by \textit{s}-\textit{d} orbital hybridization in
  heavy metals {\em Phys. Rev. B\/} {\bf 105}(21) 214438
  \urlprefix\url{https://doi.org/10.1103/PhysRevB.105.214438}

\bibitem{Shiomi2018}
Shiomi Y, Lustikova J, Watanabe S, Hirobe D, Takahashi S and Saitoh E 2018 Spin
  pumping from nuclear spin waves {\em Nature Physics\/} {\bf 15} 22--26
  \urlprefix\url{https://doi.org/10.1038/s41567-018-0310-x}

\bibitem{Kikkawa2021}
Kikkawa T, Reitz D, Ito H, Makiuchi T, Sugimoto T, Tsunekawa K, Daimon S,
  Oyanagi K, Ramos R, Takahashi S, Shiomi Y, Tserkovnyak Y and Saitoh E 2021
  Observation of nuclear-spin seebeck effect {\em Nature Communications\/} {\bf
  12} \urlprefix\url{https://doi.org/10.1038/s41467-021-24623-6}

\bibitem{Cahaya2021}
Cahaya A~B 2021 Antiferromagnetic spin pumping via hyperfine interaction {\em
  Hyperfine Interactions\/} {\bf 242} 46 ISSN 1572-9540
  \urlprefix\url{https://doi.org/10.1007/s10751-021-01780-0}

\bibitem{10.1143/PTP.101.11}
Akai H, Akai M, Blügel S, Drittler B, Ebert H, Terakura K, Zeller R and
  Dederichs P~H 1990 {Theory of Hyperfine Interactions in Metals} {\em Progress
  of Theoretical Physics Supplement\/} {\bf 101} 11--77 ISSN 0375-9687
  (\textit{Preprint} \eprint{https://doi.org/10.1143/PTP.101.11})
  \urlprefix\url{https://doi.org/10.1143/PTP.101.11}

\bibitem{book:1567088}
Freeman A~J and Watson R~E 1965 Hyperfine interactions in magnetic materials
  {\em Magnetism, Vol. 2A: Statistical Models, Magnetic Symmetry, Hyperfine
  Interactions, and Metals\/} ed Rado G and Suhl H (Academic Press Inc) pp
  168--305

\bibitem{Kubo}
Kubo R 1957 Statistical-mechanical theory of irreversible processes. i. general
  theory and simple applications to magnetic and conduction problems {\em
  Journal of the Physical Society of Japan\/} {\bf 12} 570--586
  (\textit{Preprint} \eprint{https://doi.org/10.1143/JPSJ.12.570})
  \urlprefix\url{https://doi.org/10.1143/JPSJ.12.570}

\bibitem{PhysRevA.10.2461}
Visscher W~M 1974 Transport processes in solids and linear-response theory {\em
  Phys. Rev. A\/} {\bf 10}(6) 2461--2472
  \urlprefix\url{https://doi.org/10.1103/PhysRevA.10.2461}

\bibitem{CAHAYA2022168874}
Cahaya A~B 2022 Adiabatic limit of rkky range function in one dimension {\em
  Journal of Magnetism and Magnetic Materials\/} {\bf 547} 168874 ISSN
  0304-8853 \urlprefix\url{https://doi.org/10.1016/j.jmmm.2021.168874}

\bibitem{KITTEL19691}
Kittel C 1969 Indirect exchange interactions in metals {\em Solid State
  Physics\/} vol~22 ed Seitz F, Turnbull D and Ehrenreich H (Academic Press) pp
  1--26 \urlprefix\url{https://doi.org/10.1016/S0081-1947(08)60030-2}

\bibitem{Kim1999}
Kim D~J 1999 Linear responses of metallic electrons {\em New Perspectives in
  Magnetism of Metals\/} (Springer {US}) pp 85--146

\bibitem{Suhl}
Suhl H 1958 Effective nuclear spin interactions in ferromagnets {\em Phys.
  Rev.\/} {\bf 109}(2) 606--606
  \urlprefix\url{https://link.aps.org/doi/10.1103/PhysRev.109.606}

\bibitem{korringa1950nuclear}
Korringa J 1950 Nuclear magnetic relaxation and resonnance line shift in metals
  {\em Physica\/} {\bf 16} 601--610

\bibitem{Vagner2003}
Vagner I~D 2003 {\em Nuclear Spintronics\/} (Dordrecht: Springer Netherlands)
  pp 289--307 ISBN 978-94-010-0221-9
  \urlprefix\url{https://doi.org/10.1007/978-94-010-0221-9_23}

\bibitem{Kutzelnigg1988}
Kutzelnigg W 1988 Origin and meaning of the fermi contact interaction {\em
  Theoretica Chimica Acta\/} {\bf 73} 173--200
  \urlprefix\url{https://doi.org/10.1007/bf00528203}

\bibitem{YENSEN1928503}
Yensen T 1928 What is the magnetic permeability of iron? {\em Journal of the
  Franklin Institute\/} {\bf 206} 503--510 ISSN 0016-0032
  \urlprefix\url{https://doi.org/10.1016/S0016-0032(28)90558-X}

\bibitem{Schenck1996}
Schenck J~F 1996 The role of magnetic susceptibility in magnetic resonance
  imaging: {MRI} magnetic compatibility of the first and second kinds {\em
  Medical Physics\/} {\bf 23} 815--850
  \urlprefix\url{https://doi.org/10.1118/1.597854}

\bibitem{Bertotti2009}
Bertotti G, Mayergoyz I~D and Serpico C 2009 Nonlinear magnetization dynamics
  in nanosystems {\em Nonlinear Magnetization Dynamics in Nanosystems\/}

\bibitem{Kittel2018}
Kittel C 2018 {\em Introduction to solid state physics\/} 8th ed (Wiley) ISBN
  978-1-119-45416-8
  \urlprefix\url{https://www.wiley.com/en-ie/Kittel%27s+Introduction+to+Solid+State+Physics%2C+8th+Edition%2C+Global+Edition-p-9781119454168}

\bibitem{Yusa2005}
Yusa G, Muraki K, Takashina K, Hashimoto K and Hirayama Y 2005 Controlled
  multiple quantum coherences of nuclear spins in a nanometre-scale device {\em
  Nature\/} {\bf 434} 1001--1005
  \urlprefix\url{https://doi.org/10.1038/nature03456}

\bibitem{Ota2007}
Ota T, Yusa G, Kumada N, Miyashita S, Fujisawa T and Hirayama Y 2007
  {Decoherence of nuclear spins due to dipole-dipole interactions probed by
  resistively detected nuclear magnetic resonance} {\em Applied Physics
  Letters\/} {\bf 91} 193101 ISSN 0003-6951 (\textit{Preprint}
  \eprint{https://pubs.aip.org/aip/apl/article-pdf/doi/10.1063/1.2804011/13774325/193101\_1\_online.pdf})
  \urlprefix\url{https://doi.org/10.1063/1.2804011}

\bibitem{Liu_2018}
Liu Y, Besbas J, Wang Y, He P, Chen M, Zhu D, Wu Y, Lee J~M, Wang L, Moon J,
  Koirala N, Oh S and Yang H 2018 Direct visualization of current-induced spin
  accumulation in topological insulators {\em Nature Communications\/} {\bf 9}
  \urlprefix\url{https://doi.org/10.1038%2Fs41467-018-04939-6}

\end{thebibliography}

\providecommand{\newblock}{}

\end{document}